\documentclass[12pt]{article}

\usepackage{amsmath,amssymb,amsfonts}
\usepackage[paper=letterpaper,margin=0.9in]{geometry}

\newcommand{\be}{\begin{equation}}
\newcommand{\ee}{\end{equation}}
\newcommand{\bea}{\begin{eqnarray}}
\newcommand{\eea}{\end{eqnarray}}
\newcommand{\ba}{\begin{array}}
\newcommand{\ea}{\end{array}}

\newcommand{\eref}[1]{(\ref{#1})}

\newcommand{\CN}{{\cal N}}

\newcommand{\BC}{{\mathbb C}}
\newcommand{\CP}[1]{\BC{\mathbb P}^{#1}}

\newcommand{\vev}[1]{\langle{#1}\rangle}

\newcommand{\comment}[1]{}

\begin{document}

\rightline{{\tt hep-th/0705.4581}}
\rightline{DCPT-07/23}
\rightline{VPI-IPNAS-07-04}

\vskip 2.00 cm
\renewcommand{\thefootnote}{\fnsymbol{footnote}}
\centerline{\Large \bf Fine Structure of Dark Energy and New Physics}
\vskip 0.75 cm

\centerline{{\bf
Vishnu Jejjala,${}^{1}$\footnote{\tt vishnu.jejjala@durham.ac.uk}
Michael Kavic,${}^{2}$\footnote{\tt kavic@vt.edu} and
Djordje Minic${}^{2}$\footnote{\tt dminic@vt.edu}
}}

\vskip .5cm
\centerline{${}^1$\it Department of Mathematical Sciences,}
\centerline{\it Durham University,}
\centerline{\it South Road, Durham DH1 3LE, U.K.}
\vskip .5cm
\centerline{${}^2$\it Institute for Particle, Nuclear, and Astronomical Sciences,}
\centerline{\it Department of Physics, Virginia Tech,}
\centerline{\it Blacksburg, VA 24061, U.S.A.}
\vskip .5cm

\setcounter{footnote}{0}
\renewcommand{\thefootnote}{\arabic{footnote}}

\begin{abstract}
Following our recent work on the cosmological constant problem, in this letter we make a specific proposal regarding the fine structure ({\em i.e.}, the spectrum) of dark energy.
The proposal is motivated by a deep analogy between the blackbody radiation problem, which led to the development of quantum theory, and the cosmological constant problem, which we have recently argued calls for a conceptual extension of the quantum theory.
We argue that the fine structure of dark energy is governed by a Wien distribution, indicating its dual quantum and classical nature.
We discuss observational consequences of such a picture of dark energy and constrain the distribution function.
\end{abstract}

\newpage

\section{Dark Energy and New Physics}

Our Universe is approximately four dimensional de Sitter space with a cosmological constant $\Lambda \simeq 10^{-47}\ {\rm GeV}^4$ \cite{wmap}.
The Planck mass, $M_{\rm Pl} \simeq 10^{19}\ {\rm GeV}$, however, supplies the natural scale for a quantum theory of gravitation.
Explaining the origin of the small dimensionless number $\Lambda/M_{\rm Pl}^4$ is the cosmological constant problem \cite{revs}.
The cosmological constant problem concerns physics at both ultraviolet and infrared energy scales.
In the ultraviolet, the cosmological constant computes the energy density of the vacuum.
In the infrared, the cosmological constant determines the large-scale structure of spacetime.

Recently, we have turned the cosmological constant problem around to argue the existence of a quantum version of the equivalence principle that allows the gauging of the geometric $\CP{n}$ structure of the canonical quantum theory in much the same way that the Lorentz group is gauged to the general diffeomorphism group in going from Special to General Relativity \cite{jm}. 
Crucially, the gauging is in the configuration space of the quantum mechanics, not in spacetime.
This provides a framework for a theory of quantum gravity consistent with unitarity and the principle of holography \cite{mt,review,dj}.
Locally the physics is Matrix theory in a flat background \cite{bfss}.
The obstruction to patching the flat backgrounds together is the cosmological constant.
This is a statement of the non-decoupling of physics in the ultraviolet and the infrared in quantum gravity.

According to our proposal \cite{jm},
the vacuum energy density $\Lambda$ is dynamical and fluctuates around zero (this value is fixed by diffeomorphism invariance in the configuration space of the quantum theory).
This is to say, the cosmological constant is a random variable from the point of view of the effective classical Lagrangian.\footnote{
We adopt the perspective that although critical string theory is ten-dimensional, only four of the dimensions are large.
The details of the physics of the compact directions do not matter for present purposes.} 
In the Einstein--Hilbert action, the cosmological constant term appears as a multiplier of the volume of spacetime:
\be
S_{EH} \supset \Lambda \int d^4x\ \sqrt{-g} = \Lambda\, V.
\ee
Using the large volume approximation of the non-linear Wheeler--de-Witt equation,
we regard $\Lambda$ and $V$ as conjugate quantities that realize an uncertainty relation:
\be
\Delta\Lambda\, \Delta V \sim \hbar.
\label{unc}
\ee
% Here, $\Lambda$ is an ``energy'', while the observed volume of spacetime is ``time.''
The vacuum energy density that is measured is the fluctuation $\Delta\Lambda$ about the expected value $\Lambda = 0$.
The notion of conjugation is well defined, but approximate in our scheme.\footnote{
For a detailed discussion of the relation between $\Lambda$ and $V$ see appendix 3 of \cite{review}.
Also, for work in a similar spirit, see \cite{sork, pad, vol}.}
\comment{, as implied by the expansion about the static Fubini--Study metric.}

% \item While it is true that the uncertainty relation $\Delta\Lambda\,\Delta V\sim \hbar$ is consistent with the observed vacuum energy of our Universe, there is a problem with this approximate conjugate relation: what fixes the volume?
The smallness of the measured cosmological constant relies on the largeness of the observed spacetime.
% (This is also a problem with unimodular gravity \cite{sork}, in which there is no {\em a priori} explanation for why the Universe is big.)
We motivate the largeness of observed $V$ through a gravitational see-saw \cite{jm,jmt,jlm}.
The scale of the vacuum energy is set by the balancing of the scale of cosmological supersymmetry breaking with the Planck scale.
The UV/IR correspondence inherent to this argument depends crucially on the spacetime uncertainty relations of Matrix theory \cite{yon}.
In perturbative string theory, modular invariance on the worldsheet translates in target space to the spacetime uncertainty relation:
\be
\Delta T \, \Delta X_{\rm tr} \sim \ell_s^2 \sim \alpha'.
\ee
Here, $T$ is a timelike direction, and $X_{\rm tr}$ is a spacelike direction transverse to the lightcone.
In Matrix theory this becomes a cubic relation
\be
\Delta T \, \Delta X_{\rm tr} \, \Delta X_{\rm long} \sim \ell_{\rm Pl}^3,
\ee
where $X_{\rm long} $ is the longitudinal direction.
In the generalized quantum theory
\be
\hbar\, \Delta s \sim M_{\rm Pl}\, \Delta T.
\label{eq:deltas}
\ee
The distance $\Delta s$ on the configuration space is a real quantity --- this is true even in ordinary quantum theory --- and is proportional to the modulus of the square of the overlap between states, which is a real quantity.
This can be estimated as usual by the Euclidean path integral:
$ds \sim e^{-S_{\rm eff}}$,
where $S_{\rm eff}$ denotes a hard-to-compute-from-first-principles low-energy (Euclidean) effective action\footnote{ 
We stress that the effective action can be written in the Euclidean signature because we are considering the distances between states in the configuration space of the generalized quantum theory. 
The Lorentzian nature of the effective spacetime background comes from a particular limit used in Matrix theory to reproduce the Lorentzian asymptotic flat space.}
for the matter degrees of freedom propagating in an emergent (fixed) spacetime background, we obtain a gravitational see-saw formula
\begin{equation}
\Delta X_{\rm tr} \, \Delta X_{\rm long} \sim e^{S_{\rm eff}}\, \ell_{\rm Pl}^2.
\end{equation}
The product of the ultraviolet cutoff (the maximal uncertainty in the transverse coordinate) and the infrared cutoff (the maximal uncertainty in the longitudinal coordinate) is thus exponentially suppressed compared to the Planck scale.
The mid-energy scale is related to a supersymmetry breaking scale.

\comment{
Even though, the breaking of supersymmetry is crucial for the stability of local regions of the global spacetime manifold, in Minkowski space the cosmological constant vanishes identically.
Locally, physics is described by Matrix theory, which is a supersymmetric theory of quantum mechanics.
The fluctuations in $\Lambda$ which account for the measured vacuum energy arise as a consequence of the tension between global and local physics (UV and IR).
This is a statement about the failure of decoupling in quantum gravity.
Effective field theory, which is extraordinarily successful in its domain of validity, relies on the separation of scales, which we do not have.}

We expect that the fluctuation about the zero value is biased towards the positive sign by supersymmetry breaking.
It is therefore our generic expectation that the vacuum energy ought to scale as $m_{\rm susy}^8/M_{\rm Pl}^4$, which is consistent with the cosmology of the present de Sitter epoch.\footnote{
See also \cite{tb}.}
The considerations presented here and explored to date in our prior work \cite{jm,mt} are, however, thermodynamic in nature.
As well, a more refined statistical analysis is necessary in order for us to explore the fluctuations about $\Lambda=0$ and their possible observation.

\comment{The coincidence problem ---
why $\Omega_\Lambda \approx \Omega_{\rm matter}$ today ---
is considerably more subtle.
Weinberg's classic argument based on the Bayesian distribution of the cosmological constant and observer bias \cite{sw} may perhaps be replaced by a bias towards a certain set of observables in the proposed background independent quantum theory of gravity.
These observables would be relevant for describing the low-energy physics in which the supersymmetry breaking scale is related to the cosmological scale by the gravitational see-saw.}

In this article we consider possible effects of the new physics outlined above on the fine structure of dark energy.
In particular we argue that one can speak about the spectrum of dark energy governed by a very specific distribution which embodies both its quantum and classical aspects.
This fine structure of dark energy should, in principle, have observable effects.

\section{Blackbody Radiation and Dark Energy: An analogy}

We motivate our discussion of the spectral distribution of dark energy by an illuminating analogy with the problem of black body radiation (and specific heats) in pre-quantum physics.
In that case there is a $\frac12 k_B T$ contribution to the energy for each independent degree of freedom:
\be
dE = \sum_n \left( \frac{1}{2} k_B T \right),
\ee
where $n$ is an abstract index that labels the degrees of freedoms.
This should be compared to the cosmological constant which counts degrees of freedom in the vacuum.
Heuristically, we sum the zero-point energies of harmonic oscillators and write
\be
E_{\rm vac} = \sum_{\vec{k}} \left( \frac{1}{2}\hbar \omega_{\vec{k}} \right),
\ee
where, unlike the fixed temperature $T$, $\omega_{\vec{k}} = \sqrt{|\vec{k}|^2 + m^2}$.
The divergence of the blackbody $dE$ is the ultraviolet catastrophe that the Planck distribution remedies.
Quantum mechanics resolves the over counting.
In asking why the vacuum energy is so small, we seek to learn how quantum gravity resolves the over counting of the degrees of freedom in the ultraviolet.\footnote{
Similarly, in the infrared, the proper formulation of quantum theory of gravity should resolve the stability problem
(``Why doesn't the Universe have a Planckian size?''),
once again in analogy with the resolution of the problem of atomic stability offered by quantum mechanics.}

This analogy between blackbody and the vacuum energy problems extends even further:
\begin{itemize}
\item The total radiation density of a blackbody at a temperature $T$ is given by the Stefan--Boltzmann law:
\be
u(T) = \sigma T^4.
\ee
This is to be compared with the quartic divergence of the vacuum energy,
\be
E_{\rm vac} \sim E_0^4,
\ee
$E_0$ being the characteristic energy cut-off, for bosons, or fermions separately, up to a sign difference.
We disregard, for the moment, the cancellation that happens in supersymmetric theories which leads to a quadratic divergence.
This is appropriate in that, as noted above, in our proposal supersymmetry should be broken by new curvature effects in the generalized quantum theory, which we term cosmological breaking of supersymmetry.

\item From adiabaticity, we obtain the Wien displacement law:
\be
\omega R = {\rm constant}, \qquad
\frac{\omega}{T} = {\rm constant},
\ee
where $R$ is the size of the blackbody cavity and $\omega$ the angular frequency.
This is to be compared with the uncertainty relation \eref{unc}, which tells us that $\Delta \Lambda\, \Delta V \sim \hbar$.

More precisely, fluctuations in the volume of spacetime are fixed by statistical fluctuations in the number of degrees of freedom of the gauged quantum mechanics.
In Matrix theory, the eigenvalues of the matrices denote the positions of D$0$-branes which give rise to coherent states in gravity.
Off-diagonal terms in Matrix theory break the permutation symmetry and render the D$0$-branes distinguishable.
Therefore, to enumerate the degrees of freedom, we employ the statistics of distinguishable particles (which will be of central importance in what follows).
The fluctuation is given by a Poisson distribution, which is typical for coherent states.
The fluctuation of relevance for us is in the number of Planck sized cells that fill up the configuration space (the space in which quantum events transpire), that is to say in four-dimensional spacetime:
\be
\CN_{\rm cells} \sim \frac{V}{\ell_{\rm Pl}^4} \Longrightarrow
\Delta \CN_{\rm cells} \sim \sqrt{\CN_{\rm cells}} \Longrightarrow
\Delta V \sim \sqrt{V}\ \ell_{\rm Pl}^{2},
\ee
and thus
\be
\Delta \Lambda\, \sqrt{V}\, G_N\sim 1,
\ee
where $V$ is the observed spacetime volume and $G_N$ is the four-dimensional Newton constant \cite{jm}.
\end{itemize}

The Stefan--Boltzmann law and Wien's law are implicated in the derivation of the Planck distribution for blackbody radiation.
If the analogy holds, what does this say for vacuum energy?
A natural question to ask here is whether there is a universal energy distribution for dark energy.
If so, what is its nature and what are the observational consequences?
Here we will start with an assumption that there is such a distribution, which is natural from the point of view of the new physics advocated in the previous section.
We investigate the nature of such a distribution and consider its observational consequences.

We should note that an important consequence of this analogy is that one should compare the temperature of the cosmic microwave background radiation (CMBR) we see now, $T_{\gamma} = 2.7\ {\rm K}$, to the cosmological constant we observe now!
The spectral distribution of dark energy should then be a function of energy for the fixed present value of the cosmological constant, corresponding to the energy scale of $10^{-3}\ {\rm eV}$, in analogy with the CMBR spectral distribution.
The question of why this scale is so low (why the Universe is so big), the proposed answer to which has been outlined above, is thus analogous to the question why the background CMBR temperature is so close to the absolute zero.

\section{M-theory and Wien Distribution}

According to our proposal \cite{jm, mt, review}, M-theory is background independent Matrix theory.
The infinite momentum limit of M-theory is equivalent to the $N\to\infty$ limit of coincident D$0$-branes given by $U(N)$ super-Yang--Mills gauge theory \cite{bfss}.
In particular, Matrix theory gravitons are bound states of D$0$-branes and the gravitational interaction, and thus the geometry of spacetime, is contained in the open string dynamics, {\em viz.}\ the quantum fluctuations of matrix degrees of freedom.
D$0$-branes obey $U(\infty)$ statistics.
Infinite statistics \cite{green,st,inf,mi} can be obtained from the $q=0$ deformation of the Heisenberg algebra
\be
a_i a_j^\dagger - q a_j^\dagger a_i = \delta_{ij}, \qquad a_i|0\rangle = 0.
\ee
(The cases $q=\pm 1$ correspond to Bose and Fermi statistics; $q=0$ is the so called Cuntz algebra \cite{cuntz} corresponding to infinite statistics.)
In particular, the inner product of two $N$-particle states is
\be
\langle0| a_{i_N} \cdots a_{i_1} a_{j_1}^\dagger \cdots a_{j_N}^\dagger |0\rangle = \delta_{i_1 j_1}\cdots \delta_{i_N j_N}.
\ee
Thus any two states obtained from acting with the same creation and annihilation operators in a different order are mutually orthogonal.
The partition function is
\be
Z = \sum_{\rm states} e^{-\beta H}.
\ee
The D$0$-branes are distinguishable.
Thus there is no Gibbs factor.

Strominger has argued that charged extremal black holes obey infinite statistics \cite{st}.
Assuming that the quantum state of each black hole is a functional on the space of closed three-geometries, consider the statistics of two black holes connected by a wormhole.
Black hole exchange amounts to swapping the ends of the connecting wormhole.
In quantum gravitational systems, the wave function should be invariant under all diffeomorphisms that are asymptotically trivial and deformable to the identity.
However, the exchange of charged extremal black holes creates a different three-geometry.
This implies that the interchange is not a diffeomorphism, and the wave function is not bound by any particular symmetry property under the exchange.
Thus the wave function for many similarly charged black holes is a function of each black hole's position, and the black holes are rendered distinguishable.
This is similar to the wave function of many identical particles each with a different internal state.
Treating the black holes as particles we note that they can be in any representation of the symmetric group.
Thus we are led to conclude that these types of black holes obey infinite statistics.

This analysis is centered around the invariance of the wave function under spacetime diffeomorphisms and is carried out semiclassically.
Thus standard notions of spacetime are applicable.
In the proposed background independent Matrix theory, spacetime diffeomorphisms emerge from the diffeomorphisms of the underlying quantum phase space.
That D$0$-branes obey infinite statistics is in some sense an analogous phenomena.
In both cases the solitonic objects possess differing internal states.
This implies the exchange operator is not a diffeomorphism.
If the exchange of the D$0$-branes is not a diffeomorphism of the quantum phase space, the D$0$-branes are rendered distinguishable.
Thus they can be in any representation of the symmetric group and consequently obey infinite statistics.\footnote{
Solitonic objects in string theory do not generically obey infinite statistics.
It is the requirement of diffeomorphic invariance on the space of quantum states, whose elements are D$0$-branes, that is central to the manifestation of infinite statistics here.}

It was noted by Greenberg in \cite{green} that any theory of particles obeying infinite statistics possesses a form of non-locality.
The number operator for example is non-local and non-polynomial when written in terms of field operators:
\begin{equation}
N_{i} =a_i^{\dagger} a_i + \sum_m a_m^{\dagger} a_i^{\dagger}a_i a_m + \sum_{m_1, m_2} a_{m_1}^{\dagger} a_{m_2}^{\dagger}a_i^{\dagger} a_i a_{m_2} a_{m_1} + \ldots\quad.
\end{equation}
This non-locality does not affect the formulation of a consistent non-relativistic theory.
Cluster decomposition, the CPT theorem, and a version of Wick's theorem are still valid, and the spin statistics theorem implies that particles obeying infinite statistics can be of any spin.
A quantum theory with infinite statistics remains unitary.
However, there does not exist a consistent second quantized local field theory.
The presence of non-locality while appearing to be a liability may in fact be a virtue.
Because there is not a well-defined local field theory, effective field theory arguments will miss the possibility that dark energy is associated with quanta of infinite statistics.

Recently in \cite{ng} a holographic model of spacetime foam was considered.
It was argued that this type of spacetime foam implies the existence of a type of dark energy quanta obeying infinite statistics.
This is intriguing as this was conjectured using a different formalism from the current proposal.

If we consider the various instances in which infinite statistics play a role
({\em i.e.}, black hole physics, Matrix theory, holographic spacetime foam, as well as our formulation of a background independent Matrix theory),
we note a common feature.
In each of these the holographic principle \cite{ths} is central.
Holographic theories possess a manifestly non-local quality in that the internal degrees of freedom must know something about the boundary.
Thus the non-locality present in systems obeying infinite statistics and the non-locality present in holographic theories may be related.
(This was also argued in \cite{ng}.)
Perhaps the presence of infinite statistics in quantum gravitational systems is indicative of a holographic view of spacetime.

The central point of this letter is that
{\em the spectral distribution of dark energy that follows from infinite statistics is the familiar Wien distribution}.
First recall that in the context of the black body radiation, the Wien distribution may be expressed as
\be
\rho_W = \alpha \nu^3 e^{-\beta \frac{\nu}{T}}
\ee
where $\nu$ is the linear frequency ($\omega \equiv 2\pi \nu$).
Note that the Boltzmann factor, required by infinite statistics, is already present.
The prefactor $\alpha \nu^3$ is energy times the phase space factor, which is responsible for the compatibility of this distribution with the Stefan--Boltzmann law.
Thus a quantum Boltzmann distribution, which is what infinite statistics represents, is captured by the Wien distribution.
Recall, however, that there is a semiclassical character to the distribution function ({\em i.e.}, photons are treated as ultra-relativistic, distinguishable particles). This is, of course, nothing but the classical limit ($h\nu \gg k_B T$) of the Planck distribution.

 From here the entropy of an ideal gas governed by the Wien distribution (as well known from Einstein's pioneering paper on photons \cite{al}) is
\be
S(\nu, V, E) - S(\nu, V_0, E) = \frac{E}{\beta \nu} \ln{\frac{V}{V_0}}.
\ee
Finally the dispersion of energy is purely quantum, {\em i.e.}, particle-like, which is another crucial remark of Einstein:
\be
\langle \epsilon^2 \rangle = h \nu \rho_W d \nu.
\ee
By analogy, for the dark energy spectral function we have
\bea
\label{eq:wien}
&& \rho_{\rm DE} (E, E_0) = A E^3 e^{- B\frac{ E}{E_0}}, \\
&& \rho_{\rm vac} = \int_0^{E_0} dE\ \rho_{\rm DE}(E,E_0) \sim \frac{6 A}{B^4}\ E_0^4,
\eea
with $A,B$ universal constants, and $E_0 \sim 10^{-3}\ {\rm eV}$, which corresponds to the observed cosmological constant.
The integrated energy density is proportional to $E_0^4$, as it must be.
This Wien-like spectral distribution for dark energy is thus the central prediction of a detailed analogy between the blackbody radiation and dark energy.
This in turn is rooted in our new viewpoint on the cosmological constant problem as summarized in the introduction to this letter.
The constants $A$ and $B$ are in principle computable in the framework of the background independent Matrix theory, but that computation is forbidding at the moment.
We will therefore only concentrate on global features of this viewpoint on the fine structure of dark energy.
Also, the precise dispersion relation of the dark energy quanta
(ultimately determined by the degrees of freedom of Matrix theory within the framework of the generalized quantum theory that we have proposed)
is not relevant for the general statistical discussion of possible observational signatures presented below.

Vacuum energy ({\em i.e.}, $\sum_{\vec{k}} \frac12 \hbar\,\omega_{\vec{k}}$) has negative effective pressure.
The Wien and Planck distribution share a common prefactor, which is the reason why we argue that at low energies our proposal is consistent with the positive cosmological constant, the dark energy being modeled as vacuum energy.
 From the effective Lagrangian point of view, the positive cosmological constant accounts for the accelerated expansion.
At short distances, we have a radically different situation.
The pressure in this scenario is positive and set by the scale of $E_0$.
The proposed dark energy quanta that are physically responsible for such an effective view of the cosmological constant have a strange statistics fixed by symmetry requirements, and which has certain parameters that should be bounded by observation.\footnote{
A useful comparison is the following.
For photons in the CMBR there exists a vacuum contribution and then the usual Planck distribution.
Ours is a completely analogous claim:
we have the vacuum part and the distribution of the quanta which constitute the vacuum.
The only difference here is that the quanta are unusual and the distribution is unusual due to the infinite statistics invoked.}

To summarize,
in accordance with our view of the cosmological constant problem, we think of dark energy as vacuum energy.
Just as in the case of a photon gas, the Wien distribution for vacuum energy exhibits both a classical and a quantum nature.
In Matrix theory the degrees of freedom, in the infinite momentum frame, are non-relativistic and distinguishable D$0$-branes whose dynamics are obtained from a matrix quantum mechanics.
The UV/IR correspondence at the heart of Matrix theory (and holographic theories in general) encodes the essential dualism of the cosmological constant problem:
vacuum degrees of freedom determine the large-scale structure of spacetime.

With this in mind,
the natural question to ask about this hypothesis is the following.
Can this dark energy Wien distribution, or its other consequences be directly observed?\footnote{
We thank Nemanja Kaloper for characteristically incisive questions and a very generous sharing of information pertaining to this crucial issue.}

\section{Possible Observational Consequences}

Direct observation of the Wien distribution for dark energy from calorimetry, {\em i.e.}, the analogue of measurements of the CMBR, is probably impossible,
given the gravitational nature of Matrix theory degrees of freedom.
We mention some more practical tests that one might be able to make of our proposal.

\begin{itemize}
\item Recently, a possibility for a direct observation of dark energy in the laboratory has been discussed in the literature \cite{beck,copeland}.
The idea is simple and fascinating.
One simply relies on identifying dark energy as the quantum noise of the vacuum, as governed by the fluctuation-dissipation theorem.
For example, by assuming that vacuum fluctuations are electromagnetic in nature, the zero point energy density is given by the phase space factor of the Planck distribution
(the same as the one discussed above in the case of the Wien distribution).
The integrated expression, which formally diverges, if cut-off by the observed value of dark energy, $E_0$, would correspond to the cut-off frequency
\be
\nu_{\rm DE} \sim 1.7 \times 10^{12}\ {\rm Hz}.
\ee
The present experimental bound \cite{beck,copeland} is around $\nu_{\rm max}\sim 6 \times 10^{12}\ {\rm Hz}$.

If our proposal is correct, and the dark energy is endowed with its own spectral distribution of the Wien type, then there is a window around the $\nu_{\rm DE}$ determined by the fluctuations $\delta E_0$ of dark energy around $E_0$.
The present maximum frequency can be viewed as a bound on the possible fluctuation $\delta E_0$.
The theoretical value of this fluctuation is tied to the precise value of the parameters in the Wien distribution, which are determined by the underlying new physics.

The fluctuation in the dark energy distribution (\ref{eq:wien}) is
\be
\frac{\delta\rho_{\rm DE}}{\rho_{\rm DE}} = \frac{B E}{E_0^2}\ \delta E_0.
\ee
We have as well
\be
\delta E^2 = \vev{E^2} - \vev{E}^2 = \frac{4 E_0^2}{B^2},
\ee
where
\be
\vev{E^a} = \frac{\int_0^\infty dE\ E^a\ \rho_{\rm DE}(E,E_0)}{\int_0^\infty dE\ \rho_{\rm DE}(E,E_0)}.
\ee

The observed vacuum energy is given as
\be
\int_{0}^{\nu_{\rm DE}} d\nu\ \rho_{\nu} = \frac{\pi h}{c^3} \nu_{\rm  DE}^4.
\ee
Now, we identify $\delta E$ with the fluctuation of the vacuum energy around $E_0$.
The energy density corresponding to the maximum observed frequency should bound the fluctuation of $E_0$.
This implies
\be
\delta E = \delta E_0 = \frac{2 E_0}{B} \leq E_0\left(1-\frac{\nu_{\rm max}}{\nu_{\rm DE}}\right).
\ee
Inserting the current observational bound, $\nu_{\rm max}$ and the values for $E_0$ and $\nu_{\rm DE}$ noted above, yields the following bound on the vacuum energy fluctuation
\be
{\label{qnoise}}
\delta E_0 \lesssim 6.47\times 10^{-4}\ {\rm eV},
\ee
which in turn implies
\be
B\gtrsim 3.1\,.
\ee

\item The Greisen--Zatsepin--Kuzmin (GZK) bound provides a theoretical upper limit on the energy of cosmic rays from distant sources \cite{gzk}.
In the usual GZK setup a CMBR photon is scattered off a proton producing positively charged or neutral pions (plus a neutron or a proton), thus degrading the incoming proton's energy.
The rough estimate of the energy cutoff is the threshold when the final products are both at rest.
Neglecting the split between proton and neutron masses one gets from simple kinematics
\be
E_{\rm threshold} \sim \frac{(m_p + m_{\pi})^2 - m_p^2}{4 E_{\gamma}} \sim 5\times 10^{19}\ {\rm eV}.
\ee
Note, $E_{\gamma} \sim 6.4 \times 10^{-4}\ {\rm eV}$, from the temperature of $T_{\gamma} = 2.7\ {\rm K}$,
and there are on average in one ${\rm cm}^3$ $400$ CMBR photons.
This depletion occurs on distances of $O(10)\ {\rm Mpc}$. Recently, the GZK cutoff was observed by the
Pierre Auger Observatory \cite{auger} which found a suppression in the cosmic ray spectrum above $10^{19.6}\ {\rm eV}$
at six sigma confidence.

We now consider the interaction of high energy cosmic rays with the proposed
dark energy distribution for which there should be an analogous GZK effect.
Although the coupling for the interaction responsible for this effect would be quite
small, over cosmological distances
the effect could be observable.
In our case the modification of the corresponding GZK formula, comes from a simple replacement 
of $E_{\gamma}$ by $E_0 + \delta E$, which implies
\begin{equation}
E_{\rm threshold}\simeq\frac{1}{4 E_{0}}\bigg[(m_p+m_{\pi})^2-m_p^2 - \frac{\delta E}{E_0} \Big( (m_p + m_{\pi})^2 - m_p^2 \Big)\bigg].
\label{gzk}
\end{equation}
If the fluctuation in the dark energy distribution is too great the analogous  
GZK cutoff considered here would fall below that of the standard cutoff 
and would be observed as an unexplained suppression in the cosmic ray spectrum.
No such suppression has been detected. Thus we may use the observed cosmic ray 
spectrum to further constrain the fluctuation in the dark energy distribution. 
Taking as our lower bound the observed standard GZK cutoff and making use
of (\ref{gzk}) we find 
\be
\delta E \lesssim 4.37\times 10^{-4}\ {\rm eV}.
\ee
This is a similar but more stringent bound than the one provided by quantum noise measurements, 
(\ref{qnoise}). It is worth noting that these two bounds were derived from
unrelated physical phenomena but are of the same order of magnitude. This suggests a
level of consistency in the proposal for dark energy quanta presented above.

\end{itemize}

We will briefly make note of other possible observational consequences of a distribution for the dark energy in the CMBR.
The Sunyaev--Zel'dovich (SZ) effect \cite{sz} is a combination of thermal, kinematic, and polarization effects that distort the CMBR spectrum.
The effect is an inverse Compton scattering process that serves to decrease the intensity of the Rayleigh--Jeans part of the spectrum by shifting it to higher frequency and to increase the intensity of the Wien part.
It is crucial that the effect is redshift independent.
The natural questions, in our context, are:
Are there consequences of the SZ effect if dark energy has a spectral distribution?
Does the redshift independence still apply with a distribution?

Similarly, the Sachs--Wolfe effect \cite{sachs} correlates anisotropies in the CMBR to density fluctuations.
In the case of a flat, matter dominated Friedmann--Robertson--Walker (FRW) universe, the effect of density fluctuations on the gravitational potential at the surface of last scattering is related to the temperature fluctuations by
\be
\frac{\delta T}{T} = -\frac13 \Phi.
\ee
Because the potential $\Phi$ is sensitive to the local matter density at recombination it is difficult to know how to analyze the consequence of having a distribution in the dark energy.

\section{Outlook: Dark Energy vs.\ Dark Matter}

To summarize, we have argued that dark energy has a fine structure embodied in a very particular energy distribution of a Wien type.
This distribution is compatible with the statistics of the underlying quantum gravitational degrees of freedom we have argued are relevant for a new viewpoint on the cosmological constant problem.
We have presented a preliminary discussion of possible observational implications of the dark energy spectral distribution relevant in the laboratory.

This new point of view offers other theoretical perspectives.
For example, in view of some intriguing phenomenological scaling relations found in studies of dark matter \cite{Mil, Kap},
which are apparently sensitive to the vacuum parameters, such as the cosmological constant, it is natural to ask whether within our discussion one can get both dark energy and dark matter in one go.
In Matrix theory, the open string degrees of freedom (without which we would not have infinite statistics) could thus be responsible for dark energy, and the D$0$-brane quanta attached 
to the open strings could provide natural seeds of large scale structure, {\em i.e.}, dark matter, especially when treated as non-relativistic degrees of freedom fixed to a background.
This would also imply that infinite statistics is relevant for dark matter as well!
It is intriguing that in the formal studies of infinite statistics one finds non-local expressions for the canonical fermion and boson operators in terms of Cuntz algebra ({\em i.e.}, infinite statistics) operators.
Could this mean that the standard model matter is just a collective excitation around the dark matter condensate?

Of course, such thoughts are even more speculative at this point than the argument presented in this letter.
Apart from the possible experimental tests of the dark energy spectral distribution discussed in this letter, the stringent constraints placed by the early Universe physics (for example, the details of nucleosynthesis) as well as the constraints imposed by the large scale structure (crucially dependent on dynamics of dark energy) are perhaps obvious places where further investigations of our proposal should be directed.
The interpretation of the na\"{\i}ve thermodynamic evaluation of the effective pressure, which is positive for these non-local non-linear ``quanta'', is simply an open problem, but not one that is unique to this model or excluded by data.
We intend to address these issues in future work.

\section*{Acknowledgments}
We thank many colleagues for their comments.
Our very special thanks go to Nemanja Kaloper for many discussions and constructive comments on the preliminary draft of this letter.
We have also enjoyed discussions with Vijay Balasubramanian, Jan de Boer, Eric Gimon, Rob Leigh, Tommy Levi, Jonathan Link, Bob McNees, Raju Raghavan, John Simonetti, Tatsu Takeuchi, and Chia-Hsiung Tze.
The authors acknowledge the Sowers Theoretical Physics Workshop ``What is String Theory?''\ at Virginia Tech for providing a stimulating environment for the completion of this work.
MK gratefully acknowledges Michael DuVernois, Bing Feng, Chrysostomos Dimitrios Kalousios, Manpreet Kaur, JR Newton, Bora Orcal, and Roger Wendell for their patient and thoughtful insight
and would like to especially acknowledge the shared
wisdom of Louise Marie Olsofka. MK also thanks the Theoretical Advanced Study Institute in
Elementary Particle Physics held at the University of Colorado at Boulder for hospitality during
the completion of this work.
VJ is supported by PPARC and is grateful also for the hospitality of New York University and the University of Pennsylvania.
DM is supported in part by the U.S.\ Department of Energy under contract DE-FG05-92ER40677.

\end{document}